\documentclass[sigconf,authorversion,nonacm]{acmart}

\AtBeginDocument{%
  \providecommand\BibTeX{{%
    \normalfont B\kern-0.5em{\scshape i\kern-0.25em b}\kern-0.8em\TeX}}}

\setcopyright{acmcopyright}
\copyrightyear{2022}
\acmYear{2022}
\acmDOI{XXXXXXX.XXXXXXX}
\acmConference[]{Workshop on Theories and Applications of Network Science}{December,
  2022}{Montreal, QC}
\acmPrice{15.00}
\acmISBN{978-1-4503-XXXX-X/18/06}

\usepackage{todonotes}
\usepackage{booktabs}

\makeatletter
\def\@copyrightspace{\relax}
\makeatother
\begin{document}

\title{Network Analysis of the iNaturalist\\
Citizen Science Community}

\author{Yu Lu Liu and Thomas Jiralerspong}
\affiliation{%
  \institution{McGill University}
  \city{Montreal}
  \country{Canada}}
\email{yu.l.liu@mail.mcgill.ca, thomas.jiralerspong@mail.mcgill.ca}

\renewcommand{\shortauthors}{Liu}

\begin{abstract}
In recent years, citizen science has become a larger and larger part of the scientific community. Its ability to crowd source data and expertise from thousands of citizen scientists makes it invaluable. Despite the field's growing popularity, the interactions and structure of citizen science projects are still poorly understood and under analyzed. We use the iNaturalist citizen science platform as a case study to analyze the structure of citizen science projects. We frame the data from iNaturalist as a bipartite network and use visualizations as well as established network science techniques to gain insights into the structure and interactions between users in citizen science projects. Finally, we propose a novel unique benchmark for network science research by using the iNaturalist data to create a network which has an unusual structure relative to other common benchmark networks. We demonstrate using a link prediction task that this network can be used to gain novel insights into a variety of network science methods. 
\end{abstract}
\maketitle
Note to reviewers: since our project contributes a new graph dataset, we structured the paper to be as intuitive as possible, so it does not follow the proposed structure. 
\section{Introduction and Motivation}
The field of network science and the larger scientific community have long benefited from one another. Network science methods have been used to analyze relationships in the scientific community, such as co-citation, co-authorship, etc. On the other side, datasets collected from the scientific community, for instance the citation networks Cora, Citeseer and Pubmed \cite{pubmed_cora_citeseer}, are commonly used to train and to benchmark new network science methods. Most works, however, are centered around academia: using and analysing connections between academic papers or their authors. Communities of citizen scientists are thus under-explored, both as a source of rich data and as an area of application for network science methods.

Understanding citizen science communities is also an important task in its own right. These communities contribute to science in two significant ways: i) they democratize science by involving the public in the scientific process and ii) they can be an important source of data in topics such as biodiversity, astronomy, health, etc. As an example, \textit{Bombus irisanensis Cockerell}, a vulnerable bumblebee species which has not been reported to be observed since the 1990’s, was "rediscovered" in 2020 thanks to two observations published on iNaturalist, a citizen science platform focused on gathering biodiversity organisms \cite{bumblebee}. Since citizen science platforms rely on the participation of volunteers, they face challenges related to data quality \cite{citizen_science_nature}, which is affected by the volunteers' motivation, skill and knowledge. The longevity of a platform also depends on recruitment, retention and involvement of its participants. Understanding the communities that form on these platforms are thus crucial in improving citizen science platforms, both as a data collection tool and as a social network for citizen scientists.

Our main contributions are therefore:

\begin{itemize}
\item We propose a novel collaboration network benchmark (built from a citizen science community) to the network science community. This network has an unusual structure compared to existing network science benchmarks. 
\item We demonstrate using a link prediction task that this network's unusual structure allows it to be used to gain novel insights into network science methods
\item We analyze this network using network science methods and draw insights on the current state of the citizen science community we gathered data from. 
\end{itemize}

\section{Related Work}
\label{sec:prev_work}
There have been previous papers analyzing citizen science, but they do not involve network science. For example, Aceves-Bueno et al. \cite{aceves2017accuracy} used statistical methods to analyze the accuracy of citizen science data used in various publications. Bonnet et al. \cite{bonney2016can} used statistical methods to quantify the impact that citizen science has had on scientific research, on the general public's understanding of various scientific domains, and on the personal well-being of people contributing to citizen science projects. Other studies have specifically studied how citizen scientists contribute to the iNaturalist platform using other techniques than network science: Aristeidou et al. \cite{aristeidou2021exploring} used various statistical approaches to analyze the behavior of young people contributing to the iNaturalist platform.  Di Cecco et al. \cite{di2021observing} used various statistical and clustering techniques to understand biases in citizen science data and to understand which types of users use the iNaturalist platform. Our work obviously differs from these because we propose to analyze citizen science communities under the perspective of network science.

Finally, although combining network science and citizen science is a very under-explored area, there have been a few previous studies in the area.  Pelacho et al. \cite{citizen_science}, for instance, used network science techniques to analyze the evolution and collaboration networks of scientific publications stemming from citizen science projects. Although this analysis is related and applied to citizen science, it is still in the context of academic publications and thus does not capture the full breadth of possibilities provided by citizen scientists. Mazumdar et al. \cite{fi12120210} is an another example: they applied network science techniques in order to analyze how people discuss citizen science on social media platforms such as Twitter. This application of network science, while related to citizen science, is not directly applied to citizen science communities but instead to the way people communicate about citizen science.

To the best of our knowledge, our work is the first to use network science techniques to analyze and understand the interactions within citizen science communities in a non-academic setting, to view data from the iNaturalist platform as a network, and to apply network science techniques to the iNaturalist platform.

\section{Problem Definition}
Our project focuses on citizen science communities on iNaturalist, a social network platform where users share and map observations of nature. It is a joint initiative of the California Academy of Sciences and the National Geographic Society, launched in 2008. As of 2022, the number of observations on the platform has reached 115 million \cite{wiki:INaturalist}. Users on the platform can upload pictures of organisms they observe and identify organisms in pictures that other users upload. Each observation can only reach "research grade" quality when there is a community consensus on the species of the observed organism.

The first goal of our project is to gain insights about the communities on the iNaturalist citizen science community:
\begin{itemize}
    \item Are there clusters of users who tend identify the same groups of species (e.g. birdwatchers)? 
    \item What are the species that most often identified? Does their taxonomic class come into play?
    \item How do new users (e.g. users who created their accounts in 2022) fit into the network? 
    \item etc.
\end{itemize}

To do this, in Section \ref{sec:analysis-inaturalist}, we visualize and analyze the iNaturalist network to identify interesting patterns that can be found in it.

The second goal of our project is to contribute a new benchmark network to collaboration network research. To do this, in Section \ref{sec:collaboration} we analyze iNaturalist by viewing each taxon as a "project" that users can contribute to by correctly identifying observations of said taxon. Users that identify a common taxon can thus be viewed as "collaborators", making the resulting network a collaboration network. 

In Section \ref{sec:collaboration-comparison} we first investigate whether our "collaboration" network, which is built from a citizen science platform, has a different structure than existing collaboration networks, which stem from academia. To do so, we compute network measures commonly used to analyze co-authorship networks and offer a comparison and an interpretation of the results.

In Section \ref{sec:link_pred}, we conduct further experiments to see if our network can provide valuable insights into existing network science methods by checking whether common link prediction methods perform differently or unusually on our network compared to existing collaborating networks.
\section{Data Collection}
\begin{figure}[h]
    \centering
    \includegraphics[width=0.35\textwidth]{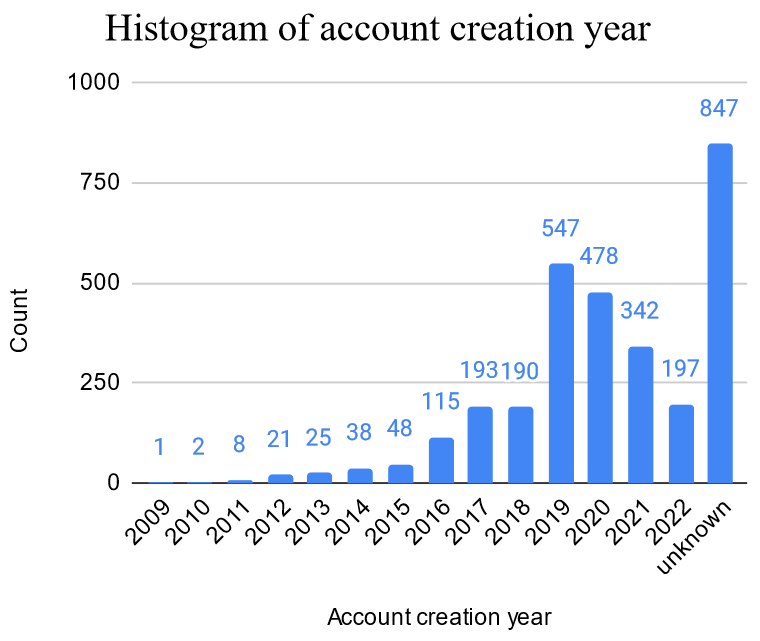}
    \caption{Histogram of users' account creation years. We see this information is unavailable for around 30\% of users.}
    \label{fig:histogram_years}
\end{figure}
To build the iNaturalist collaboration network, we start by collecting observations. We collected all research-grade observations made in the Alberta 15th Division, a region situated within the Rocky Mountains in Western Canada and includes the well-known Jasper and Banff national parks. The observations are collected via GBIF \footnote{ GBIF.org (27 November 2022) GBIF Occurrence Download https://doi.org/10.15468/dl.wt7szw }. This corresponds to a collection of 41K observations identified by 2998 users and covering 1816 species. For each observation, we log the species of the observed organism (as the corresponding taxon ID) and the ID of the user who correctly identified it.

We define two types of nodes: a user node for each distinct user ID, and a species node for each distinct taxon ID. We draw a link between a user node and a species node if the user in question has correctly identified the species at least once within our collected observations. We denote this initial bipartite graph by \textbf{\texttt{user-taxon}}. By projecting the \texttt{user-taxon} network onto the user nodes, we obtain the \textbf{\texttt{user-user}} network which contains only user nodes. Two users are connected if they have "collaborated", i.e. identified at least one common species within our collected observations. By projecting the \texttt{user-taxon} network onto the taxon nodes, we obtain the \textbf{\texttt{taxon-taxon}} network which contains only taxon nodes. Two taxons are connected if they share at least one common user-identifier within our collected observations.
\begin{figure*}[h!]
    \centering
    \includegraphics[width=1\textwidth]{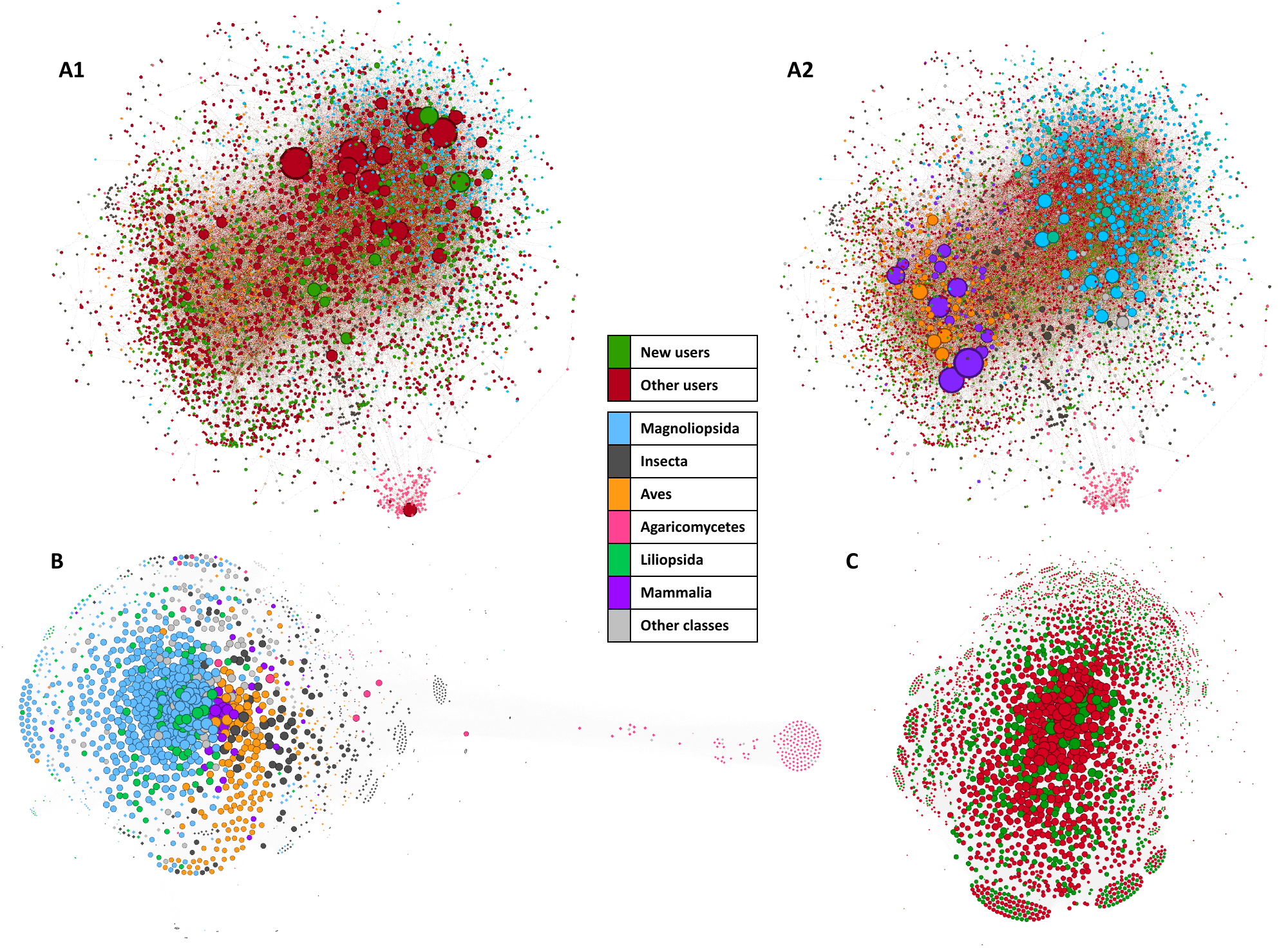}
    \caption{In A1 and A2, we visualize the bipartite graph \texttt{user-taxon}. In A1, user nodes' size is proportional to their (out) degree. In A2, taxon nodes' size is proportional to their (in) degree. In B, we visualize the \texttt{taxon-taxon} projection. In C, we visualize the \texttt{user-user} projection. User nodes' colour indicates whether they are new (joined on or after 2020). Taxon nodes' colour indicates their taxonomic class.}
    \label{fig:visualization}
\end{figure*}
We add the following additional node attributes:
\begin{itemize}
    \item Taxonomic class (e.g. insects, birds, etc.) of each taxon to taxon nodes. This additional information allows us to view similar species as a group. 
    \item Account creation year of each user to user nodes. This additional information allows us to track new users and compare their involvement within the community to older user accounts. 
\end{itemize}
We note one limitation of our approach: the account creation year is an unavailable information for some user accounts. This limits our capability to differentiate between new users and old users, so we instead analyze users known to be new (referred to as "new users" from now on) versus other users. We present a histogram of account creation years in Figure~\ref{fig:histogram_years}. We define new users to be users with accounts created on or after 2020, which corresponds to around 50\% of the user base with a known account creation year.

\section{Analysis of the iNaturalist Community} \label{sec:analysis-inaturalist}
\begin{figure*}[h]
    \centering
    \includegraphics[width=0.85\textwidth]{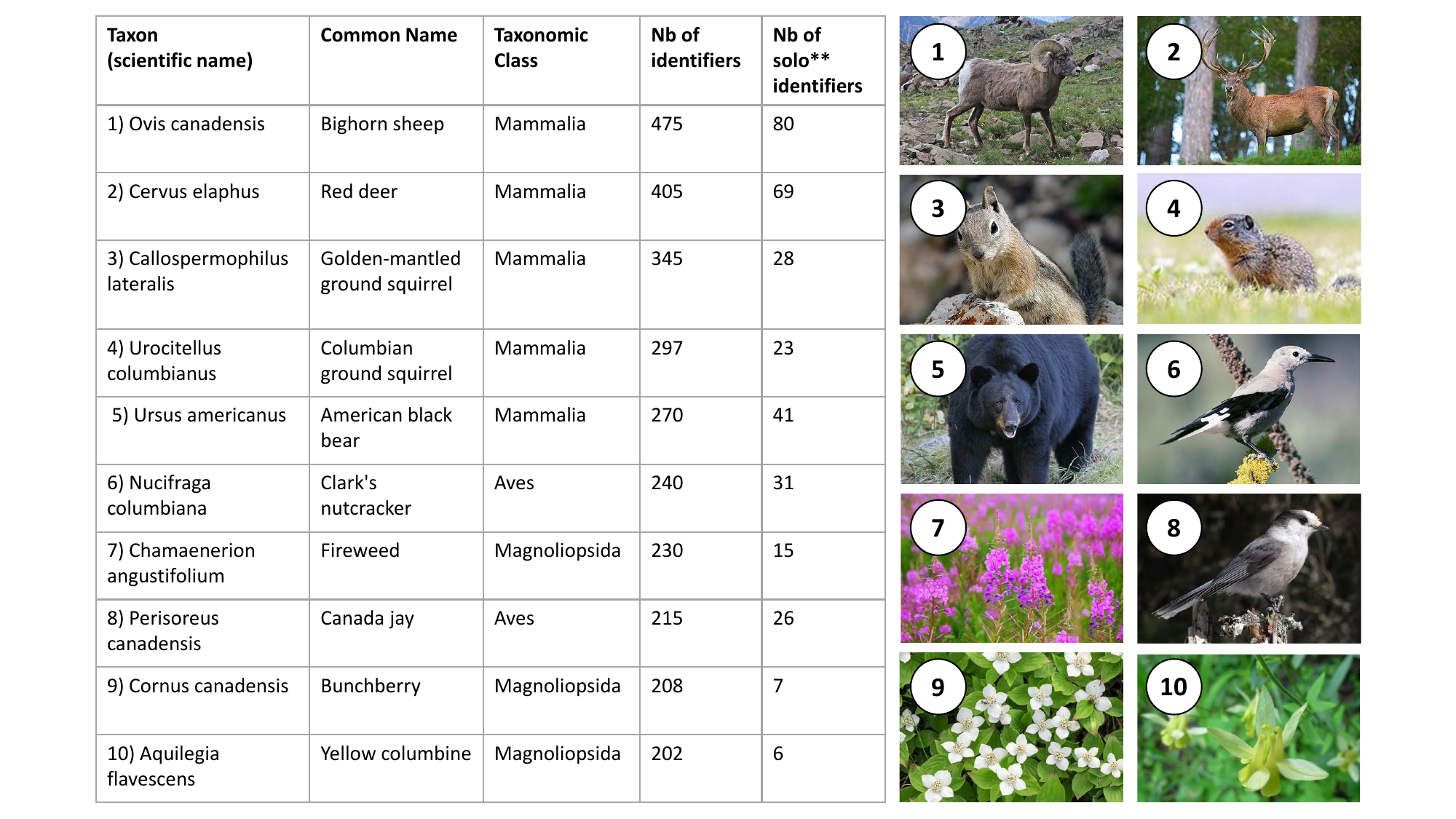}
    \caption{The top 10 taxons with the highest number of user-identifiers (in-degree). **We use "solo identifiers" to refer to users who only identified the taxon in question, within our collected observations. For instance, there are 80 users who identified bighorn sheeps and no other taxons. Source of pictures: Wikipedia \cite{wiki:Bighorn_sheep, wiki:Clark's_nutcracker, wiki:Columbian_ground_squirrel, wiki:Golden-mantled_ground_squirrel, wiki:Aquilegia_flavescens, wiki:American_black_bear, wiki:Canada_jay, wiki:Chamaenerion_angustifolium, wiki:Red_deer}}
    \label{fig:central_nodes}
\end{figure*}
We present the visualizations of three networks in Figure~\ref{fig:visualization}: the bipartite graph, the \texttt{taxon-taxon} projection and the \texttt{user-user} projection. The visualization is produced using the Force Atlas algorithm via Gephi \cite{Gephi}. Each user node is colour-coded to indicate whether the user is new (joined after 2020) or not. Each taxon node is colour-coded to indicate the taxonomic class to which it belongs:
\begin{itemize}
    \item Magnoliopsida, a class of flowering plants \cite{wiki:Magnoliopsida}: 575 taxons;
    \item Insecta, insects: 489 taxons;
    \item Aves, birds \cite{wiki:Bird}: 215 taxons;
    \item Agaricomycetes, a class of fungi which includes mushroom-forming fungi \cite{wiki:Agaricomycetes}: 180 taxons;
    \item Liliopsida, lilies\cite{wiki:Liliopsida}: 125 taxons; 
    \item Mammalia, mammals: 53 taxons;
    \item Other, taxonomic classes not included above: 179 taxons.
\end{itemize}
Our main goal in this section is to draw insights about the community using our networks.
\subsection{\texttt{user-taxon} bipartite graph}
\label{sec:user-taxon}
In the bipartite graph (Figure \ref{fig:visualization}: A1 and A2), we set edges to be directed from users to taxons so that in-degree and out-degree values can better distinguish between important user nodes and important taxon nodes. 

First, we notice an obvious but small cluster at the bottom of the graph consisting of an user who identified several Agaricomycetes. This user and this taxonomic class seem to have little "interaction" with the rest of the network. If we exclude this small cluster, the bipartite graph has a peanut-like shape: the left is largely occupied by birds, mammals and users who identify species of those two classes; the right, on the other hand, is largely occupied by taxons of the Magnoliopsida class and users who identify these flowers. This suggests that \textbf{the INaturalist community in the Alberta 15th Division can be viewed as one "specialist" individual and two groups: one focused on birds and mammals, and the other focused on Magnoliopsida flowers}.

We can gain further insights into the community by visualizing the degree values. In the first view of the graph (Figure~\ref{fig:visualization}: A1), the user node size is proportional to the node's out-degree value. In other words, big user nodes in the visualization correspond to users who identified a larger amount of distinct taxons. We can see that the biggest nodes are not new users and are placed either within the right cluster or in the middle of the two main clusters. This suggests that \textbf{users who can identify a larger amount of species tend not to be new users and that their focus is oriented more towards Magnoliopsida flowers}. Intuitively, this makes sense as Magnoliopsida is the class containing the most taxons.

In the second view of the graph (Figure~\ref{fig:visualization}: A2), the taxon node size is proportional to the node's in-degree value. This means that a taxon node is big if it has a large number of users who identified it. We can see that although there are few mammal nodes, they are the biggest: there are more users who identified mammals than taxons belonging to other taxonomic classes. Moreover, we can observe that the left cluster tends to contain larger nodes than the right cluster. This suggests that \textbf{while there are fewer species of mammals and birds, they are more easily identified by members of the community. On the other hand, while there are many species of Magnoliopsida flowers, they are in general harder to identify and are therefore able to be identified by fewer users}.

To discern the exact species that are identified by most users, we present in Figure~\ref{fig:central_nodes} the top 10 taxons with the highest in-degree value. We believe that taxons on this list are either easily identifiable/observable (more observations allow for more identifications), more interesting to identify/observe, or very likely a combination of all these factors. We note in particular that four of these taxons, namely the bighorn sheep \cite{wiki:Bighorn_sheep}, the golden-mantled ground squirrel \cite{wiki:Golden-mantled_ground_squirrel}, the Columbian ground squirrel \cite{wiki:Columbian_ground_squirrel}, and the Clark's nutcracker \cite{wiki:Clark's_nutcracker}, have the Western North American mountains as their one and only habitat, which could explain their importance within the iNaturalist community of the Alberta 15th Division.
\begin{figure*}[h!]
    \centering
    \includegraphics[width=0.85\textwidth]{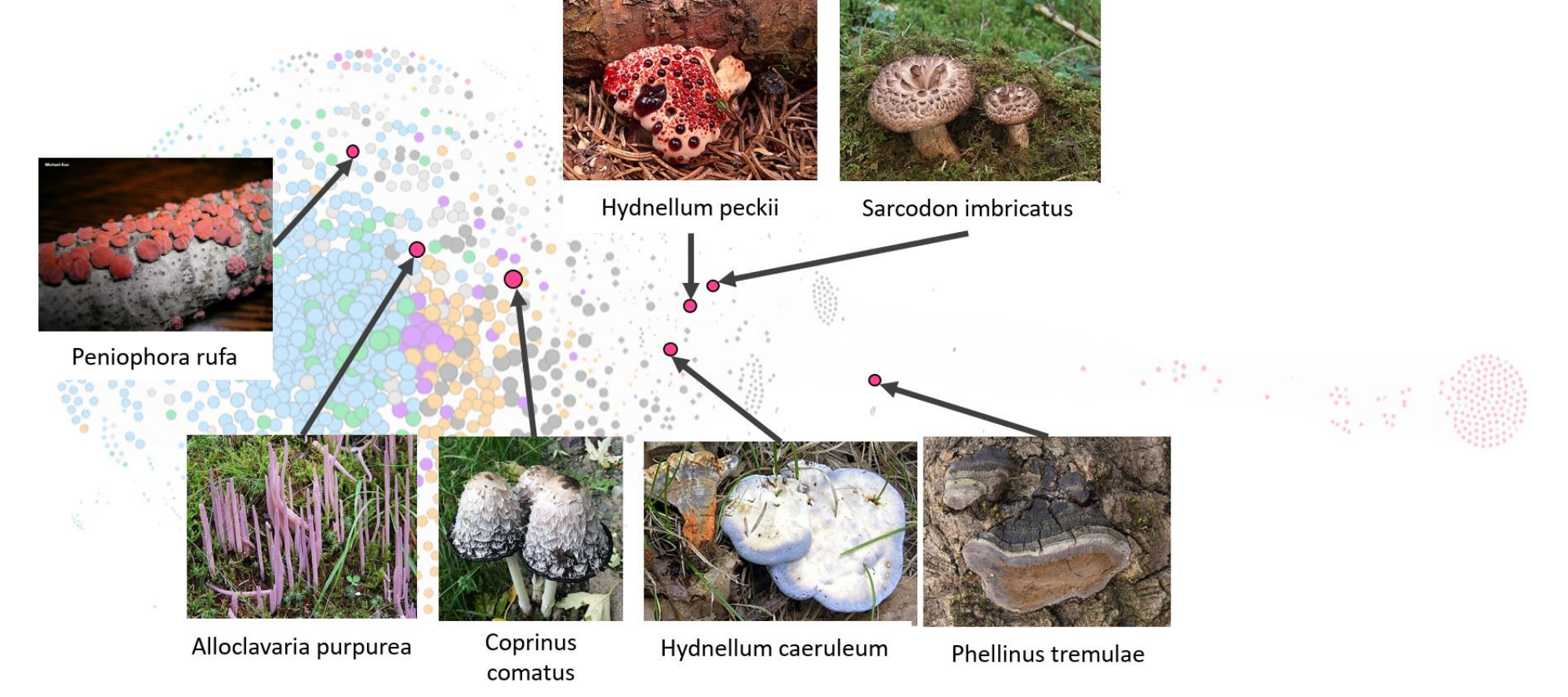}
    \caption{Agaricomycetes taxons of interest. We note that most of these taxons have distinctive colours or shape, which might make them easier or more interesting to observe/identify. Source of pictures: Wikipedia \cite{wiki:Hydnellum_pecki, wiki:Sarcodon_imbricatus, wiki:Alloclavaria_purpurea, wiki:Coprinus_comatus, wiki:Hydnellum_caeruleum, wiki:Phellinus_tremulae}}
    \label{fig:case_study_mushroom}
\end{figure*}
We also highlight a group of users we call "solo identifiers", which are users who only identified one taxon and no other taxons. If we sum up the number of solo identifiers for the top 10 taxons presented in Figure~\ref{fig:central_nodes}, we obtain a total of 326 such users, which corresponds to around 10\% of the whole user base. We elaborate on this group of users in Section~\ref{sec: user-user}.
\subsection{\texttt{taxon-taxon} projection}
\label{sec: taxon-taxon}
In the \texttt{taxon-taxon} projection (Figure~\ref{fig:visualization}: B), the node size is proportional to the node's degree, which means that bigger nodes represent taxons that share user-identifiers with a greater number of other taxons. Through this lense, we can see that the majority of bigger nodes are clustered at the center, surrounded by small clusters made of smaller nodes. By using the colour coding, we can also observe that taxons of the same taxonomic class tend to be grouped together, the most obvious cases being Agaricomycetes, Magnoliopsida, and Aves (birds). This means that \textbf{users tend to identify taxons belonging to the same taxonomic class}. For example, a user who identified a bird taxon might also identify other bird taxons, which makes all bird taxons be grouped closely together. Intuitively, this makes sense as nature enthusiasts tend to be grouped as bird-watchers, mycophiles, etc. 

Interestingly, we see bigger Agaricomycetes nodes that lie either within the main cluster or between the main cluster and the far-away Agaricomycetes cluster. We present their scientific name and their physical appearance in Figure ~\ref{fig:case_study_mushroom}. Compared to Agaricomycetes nodes in the far-away cluster which are identified by one "specialist" individual, they might be more easily identifiable, observable or more interesting to people, such that many users who usually identify taxons of other taxonomic classes can identify them. This small case study shows that the \textbf{\texttt{taxon-taxon} projection can be useful for gaining insights about the expertise of a citizen science community}. 
\begin{figure}[h]
    \centering
    \includegraphics[width=0.4\textwidth]{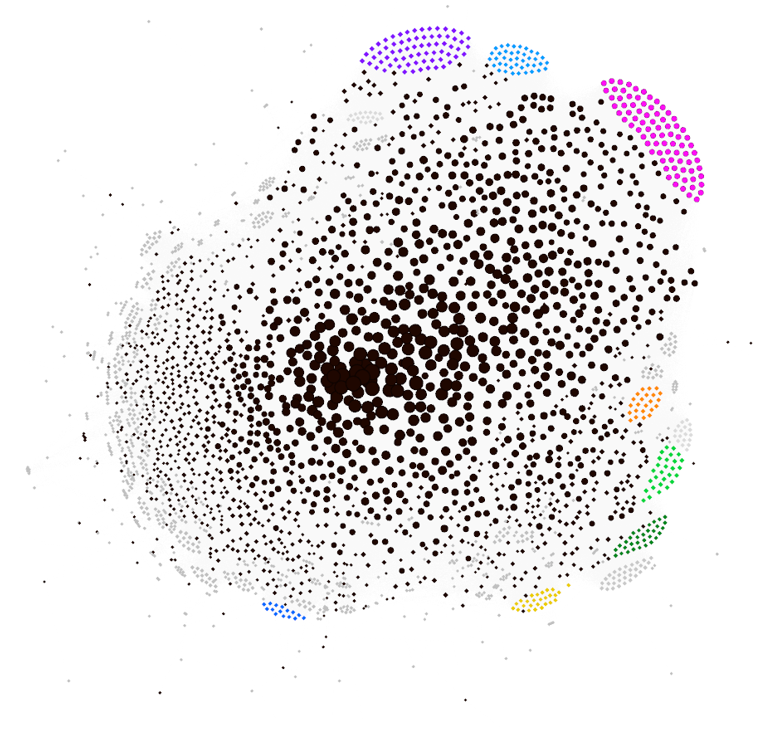}
    \caption{User-user projection with solo user-identifiers coloured in gray, solo user-identifiers who identified one of the top 10 taxons (from \ref{sec: taxon-taxon}) in bright colors. Not that only 8 visible as there are 2 taxons containing too few solo user-identifiers to be visible. Other users are coloured in black.}
    \label{fig:solo_user_visualization}
\end{figure}
\subsection{\texttt{user-user} projection}
\label{sec: user-user}
In the \texttt{user-user} projection (Figure~\ref{fig:visualization}: C), we also set the node size to be proportional to the node's degree. In this case, bigger nodes represent users who share a greater number of "projects" with other users, each "project" being the identification of a taxon. We see that bigger nodes (more collaborative users) are mostly in the middle while smaller nodes (less collaborative users) are pushed to the periphery and form small clusters. To further investigate these clusters, we return to the "solo identifiers" introduced in Section~\ref{sec:user-taxon}. In Figure~\ref{fig:solo_user_visualization}, we colour-code the solo identifiers so that their user nodes are either black, gray, or brightly coloured. Each bright colour denotes solo-identifiers of one of the top 10 taxons listed in Section~\ref{sec:user-taxon}, whereas gray denotes solo-identifiers of other taxons and black denotes other users. We see that each small cluster corresponds exactly to the solo-identifiers of one taxon and that all solo-identifier nodes are pushed to the periphery. This shows that the visualization of the \texttt{user-user} projection can be used to \textbf{identify groups of users who behave in a significantly different way than others}.

In Figure~\ref{fig:visualization} (C) we also colour-code the new users, but we do not see any obvious trend.
\begin{table*}
\begin{tabular}{ |p{4cm}||p{2cm}|p{2cm}|p{2cm}| p{2cm} |p{2cm}| }
 \hline
 & iNaturalist (ours) & GR-QC & Erdos & NetScience & CS-PhD \\
 \hline
 \hline
Number of nodes & 2998 & 4158 & 5094 & 379 & 1882\\
 \hline
Number of edges & \textbf{585115} & 13422 & 7515 & 914 & 1740\\
\hline
Number of connected components & 3 & 1 & 17 & 1 & 168\\
\hline
Size of largest component& 2994 & 4158 & 4991 & 379 & 1025\\
 \hline
Power law fit slope& -0.475 & -1.974 & -1.678 & -1.553&-2.136\\

 \hline
Average shortest path length& \textbf{1.906} & 6.049 & 5.511 & 6.042&11.659\\
\hline
Average clustering coefficient& 0.806 & 0.557 & 0.0818 & 0.741&0.00507\\
\hline
Spectral gap& \textbf{3818.504} & 34.999 & 28.143 & 6.431&5.969\\
\hline 
Degree vs. degree correlation coefficient& -0.27 & 0.639 & -0.444 & -0.0817&-0.197\\
\hline 
Degree vs. clustering coefficient correlation coefficient& -0.745 & 0.0513 & 0.0153 & -0.362&0.0195\\
\hline
\end{tabular}

\caption{Basic metrics for all collaboration networks. Outliers for the iNaturalist network are in bold}
\label{tab: basic-metrics}
\end{table*}
\section{Comparison to other collaboration networks} \label{sec:collaboration}
As mentioned previously, the \texttt{user-user} network can be viewed as a collaboration network where users are connected if they contributed to the identification "project" of a common species. We thus compare this network to existing collaboration graphs from academic scientific communities. We conduct this comparison first by computing common network metrics and comparing their values for our network and for the academic collaboration networks to see if our network has a distinct structure compared to thee academic networks. We then compare the performance of several common link prediction methods on our network compared to the academic collaboration networks to see if our network can provide interesting insights into the performance of these methods.

\subsection{Other Collaboration Networks} \label{other-networks}
We compare our user-user projection network to 4 other benchmark collaboration networks of similar size which we found in the Network Repository \cite{network_repository}:

\subsubsection*{GR-QC} \cite{leskovec_networks}

The most common form of collaboration in the academic world is co-authorship. The GR-QC collaboration network contains information about scientists who have published papers in the fields of general relativity and quantum cosmology. Each node is a scientist and there is an edge between two nodes $u$ and $v$ if scientist $u$ has co-authored a paper with scientist $v$.

\subsubsection*{Erdos} \cite{batagelj2000some}

Paul Erdos was one of the most productive mathematicians in history, publishing more than 1500 papers during his lifetime. The Erdos collaboration network is a network of Erdos, his collaborators, and his collaborators' collaborators. The network is defined in the same way as GR-QC.

\subsubsection*{NetScience} \cite{newman2006finding}

The NetScience collaboration network contains information about 379 scientists who have conducted research in the field of network science. The network is defined in the same way as GR-QC.

\subsubsection*{CS-PhD} \cite{de2011exploratory}

Although collaboration networks are often studied in the context of co-authorship, another collaborative relationship which exists in the academic world is the PhD-advisor relationship.

In the CS-PhD collaboration network, each node is a person who holds a PhD in theoretical computer science and there is an edge between two nodes $u$ and $v$ if person $u$ was the PhD advisor for person $v$.

\subsection{Basic Metrics} \label{sec:collaboration-comparison}
To compare our collaboration network to other academic collaboration networks, we first compute some basic metrics for our network as well as the benchmark networks described in section \ref{other-networks}. The results are displayed in Table \ref{tab: basic-metrics}, with outliers for the iNaturalist network in bold. We now analyze these outliers.

The iNaturalist network has a much \textbf{larger number of edges} compared to the other networks, despite having a similar number of nodes. This shows that the iNaturalist network is much denser than the academic networks, indicating that collaborations are much more frequent in citizen science compared to academic science. This is expected because participating in a citizen science taxon "project" is much more accessible than collaborating academically.

The iNaturalist network also has a much \textbf{smaller average shortest path length} compared to the academic networks. This shows that the iNaturalist network is much more small-world compared to the academic networks, which is expected due to the iNaturalist network having many more edges than the academic networks.

Finally, the iNaturalist network  has a much \textbf{larger spectral gap} compared to the other networks. A higher spectral gap indicates that a graph is more connected, so this quantitatively shows that the iNaturalist network is much more connected than the academic science networks.

These three metrics all indicate that citizen scientists tend to collaborate much more often than academic scientists. This confirms the intuitive notion that, although academic scientists may be more qualified than citizen scientists, citizen science compensates by making it easier to collaborate with others, thus allowing for a much larger number of collaborations.

We can also see from these metrics that the iNaturalist network has a very different structure compared to existing academic collaboration network benchmarks, indicating that network science algorithms may exhibit very different behavior when evaluated on the iNaturalist network compared to academic collaboration network. This might make it a valuable benchmark for network scientists to test if their algorithms are robust to many different types of network structures. We test this hypothesis using a link prediction task in the following section.

\subsection{Link Prediction in \texttt{user-user}}
\label{sec:link_pred}
\begin{table*} 
\begin{tabular}{ |p{4cm}||p{2cm}|p{2cm}|p{2cm}| p{2cm} |p{2cm}| }
 \hline
 & iNaturalist (ours) & GR-QC & Erdos & NetScience & CS-PhD \\
 \hline
 \hline
Jaccard Coefficient & 0.918 & 0.903 & 0.620 & 0.906 & 0.503\\
 \hline
Resource Allocation Index & 0.985 & 0.903 & 0.620 & 0.910 & 0.503\\
 \hline
 Adamic Adar Index & 0.960 & 0.903 & 0.620 & 0.910 & 0.503\\
 \hline
Preferential Attachment& 0.915 & 0.739 & 0.545 & 0.638 & 0.345\\
 \hline
DeepWalk& 0.843 & 0.681 & 0.857 & 0.640 & 0.647\\

 \hline
\end{tabular}

\caption{ROC-AUC scores for different link prediction methods on different collaboration network datasets}
\label{tab: link_prediction_results}
\end{table*}
We evaluate 5 link prediction methods (Jaccard Coefficient, Resource Allocation Index, Adamic Adar Index, Preferential Attachment, DeepWalk \cite{deepwalk}) on our collaboration network as well as the 4 collaboration networks from section \ref{other-networks}.
To do this we first randomly delete 20\% of the edges in the network. We call these edges "gold edges" and the edges which remain in the network "remaining edges". We then create a list of "candidate edges" by generating a complete graph on all the nodes in the network and deleting from it the edges which are part of the "remaining edges".

Finally we frame link prediction as a classification task where our link prediction method must predict whether each of the candidate edges is a gold edge or not while only having access to the remaining edges. We compute the ROC-AUC scores for each method and present the results in Table \ref{tab: link_prediction_results}.

We see that almost all the methods, with the exception of DeepWalk, perform the best on the iNaturalist network compared to the other networks, and that even DeepWalk performs second best on it. This suggests that it is easier to predict which citizen scientists have contributed to at least one of the same "projects" than it is to predict which academic scientists have co-authored a paper together. 

We also see that, for the academic networks, the 3 first methods (Jaccard Coefficient, Resource Allocation Index, Adamic Adar Index) all perform almost identically. It is thus very hard to compare these methods or say which one of them is better using these academic networks. With the iNaturalist \texttt{user-user} network, however, we see that the 3 methods' performances differ more from each other, thus allowing us to compare them and rank them. 

We also see that the 4th method (Preferential Attachment) performs much worse than the first 3 methods on the 4 academic networks, whereas it has almost identical performance to Jaccard Coefficient and only slightly worst performance than Resource Allocation Index and Adamic Adar Index on iNaturalist. This suggests that Preferential Attachment performs better on denser graphs.

DeepWalk seems to be an outlier method as its performance on the various networks does not follow the trend of the other methods: it has the worst performance on GR-QC, NetScience, and our dataset, but has the best performance on Erdos and CS-PhD. We hypothesize that this might be due to the fact that DeepWalk is the only GNN based method that we tested. GNN based methods are known to be very sensitive to the hyperparameters used to train them and we regrettably did not have time to do a proper hyperparameter search for DeepWalk. This may be why its results are outliers compared to the other methods.

These insights derived from the methods' performance on our network show that the iNaturalist \texttt{use-user} network is a valuable valuable benchmark for differentiating between methods when other benchmarks are unable to do so, and for gaining new insights into how certain methods achieve better performance on certain types of networks.

\section{Conclusion}
In this paper we used network science to gain insights from the iNaturalist community in the Alberta 15th Division. Through various visualizations and degree measures, we found that:
\begin{itemize}
    \item The INaturalist community in the Alberta 15th Division can be viewed as one individual specializing in Agaricomycetes, and two groups: one focused on birds and mammals, and the other focused on Magnoliopsida flowers.
    \item Users who can identify a larger amount of species tend not to be new users and that their focus is oriented more towards Magnoliopsida flowers.
    \item Users tend to identify taxons belonging to the same taxonomic class.
\end{itemize}

Most importantly, we demonstrated how the bipartite network and its two projection networks can be used to analyze a citizen science community.

We also introduced the iNaturalist collaboration network as a new benchmark for network science research. We showed that it has an unusual structure compared to existing collaboration network benchmarks and demonstrated using a link prediction task that this unusual structure makes it a valuable benchmark for evaluating and analyzing the properties of network science methods.

\section{Future Work}
For future work, we would like to obtain more data from the iNaturalist platform to be able to build even larger benchmark data sets. We would also like to analyze how different community detection methods perform on detecting the different taxonomic classes in the taxon-taxon projection network.

\section{Acknowledgements}
We would like to thank Professor Reihaneh Rabbany as well as all our reviewers for their valuable suggestions and feedback all throughout this project.


\bibliographystyle{ACM-Reference-Format}
\bibliography{sample-base}


\begin{thebibliography}{35}


\ifx \showCODEN    \undefined \def \showCODEN     #1{\unskip}     \fi
\ifx \showDOI      \undefined \def \showDOI       #1{#1}\fi
\ifx \showISBNx    \undefined \def \showISBNx     #1{\unskip}     \fi
\ifx \showISBNxiii \undefined \def \showISBNxiii  #1{\unskip}     \fi
\ifx \showISSN     \undefined \def \showISSN      #1{\unskip}     \fi
\ifx \showLCCN     \undefined \def \showLCCN      #1{\unskip}     \fi
\ifx \shownote     \undefined \def \shownote      #1{#1}          \fi
\ifx \showarticletitle \undefined \def \showarticletitle #1{#1}   \fi
\ifx \showURL      \undefined \def \showURL       {\relax}        \fi
\providecommand\bibfield[2]{#2}
\providecommand\bibinfo[2]{#2}
\providecommand\natexlab[1]{#1}
\providecommand\showeprint[2][]{arXiv:#2}

\bibitem[cit(2015)]%
        {citizen_science_nature}
 \bibinfo{year}{2015}\natexlab{}.
\newblock \showarticletitle{Rise of the citizen scientist}.
\newblock \bibinfo{journal}{\emph{Nature}} \bibinfo{volume}{524},
  \bibinfo{number}{7565} (\bibinfo{date}{01 Aug} \bibinfo{year}{2015}),
  \bibinfo{pages}{265--265}.
\newblock
\showISSN{1476-4687}
\urldef\tempurl%
\url{https://doi.org/10.1038/524265a}
\showDOI{\tempurl}


\bibitem[Aceves-Bueno et~al\mbox{.}(2017)]%
        {aceves2017accuracy}
\bibfield{author}{\bibinfo{person}{Er{\'e}ndira Aceves-Bueno},
  \bibinfo{person}{Adeyemi~S Adeleye}, \bibinfo{person}{Marina Feraud},
  \bibinfo{person}{Yuxiong Huang}, \bibinfo{person}{Mengya Tao},
  \bibinfo{person}{Yi Yang}, {and} \bibinfo{person}{Sarah~E Anderson}.}
  \bibinfo{year}{2017}\natexlab{}.
\newblock \showarticletitle{The accuracy of citizen science data: a
  quantitative review}.
\newblock \bibinfo{journal}{\emph{Bulletin of the Ecological Society of
  America}} \bibinfo{volume}{98}, \bibinfo{number}{4} (\bibinfo{year}{2017}),
  \bibinfo{pages}{278--290}.
\newblock


\bibitem[Aristeidou et~al\mbox{.}(2021)]%
        {aristeidou2021exploring}
\bibfield{author}{\bibinfo{person}{Maria Aristeidou},
  \bibinfo{person}{Christothea Herodotou}, \bibinfo{person}{Heidi~L Ballard},
  \bibinfo{person}{Alison~N Young}, \bibinfo{person}{Annie~E Miller},
  \bibinfo{person}{Lila Higgins}, {and} \bibinfo{person}{Rebecca~F Johnson}.}
  \bibinfo{year}{2021}\natexlab{}.
\newblock \showarticletitle{Exploring the participation of young citizen
  scientists in scientific research: The case of iNaturalist}.
\newblock \bibinfo{journal}{\emph{Plos one}} \bibinfo{volume}{16},
  \bibinfo{number}{1} (\bibinfo{year}{2021}), \bibinfo{pages}{e0245682}.
\newblock


\bibitem[Bastian et~al\mbox{.}(2009)]%
        {Gephi}
\bibfield{author}{\bibinfo{person}{Mathieu Bastian}, \bibinfo{person}{Sebastien
  Heymann}, {and} \bibinfo{person}{Mathieu Jacomy}.}
  \bibinfo{year}{2009}\natexlab{}.
\newblock \showarticletitle{Gephi: An Open Source Software for Exploring and
  Manipulating Networks}.
\newblock  (\bibinfo{year}{2009}).
\newblock
\urldef\tempurl%
\url{http://www.aaai.org/ocs/index.php/ICWSM/09/paper/view/154}
\showURL{%
\tempurl}


\bibitem[Batagelj and Mrvar(2000)]%
        {batagelj2000some}
\bibfield{author}{\bibinfo{person}{Vladimir Batagelj} {and}
  \bibinfo{person}{Andrej Mrvar}.} \bibinfo{year}{2000}\natexlab{}.
\newblock \showarticletitle{Some analyses of Erdos collaboration graph}.
\newblock \bibinfo{journal}{\emph{Social Networks}} \bibinfo{volume}{22},
  \bibinfo{number}{2} (\bibinfo{year}{2000}), \bibinfo{pages}{173--186}.
\newblock


\bibitem[Bonney et~al\mbox{.}(2016)]%
        {bonney2016can}
\bibfield{author}{\bibinfo{person}{Rick Bonney}, \bibinfo{person}{Tina~B
  Phillips}, \bibinfo{person}{Heidi~L Ballard}, {and} \bibinfo{person}{Jody~W
  Enck}.} \bibinfo{year}{2016}\natexlab{}.
\newblock \showarticletitle{Can citizen science enhance public understanding of
  science?}
\newblock \bibinfo{journal}{\emph{Public understanding of science}}
  \bibinfo{volume}{25}, \bibinfo{number}{1} (\bibinfo{year}{2016}),
  \bibinfo{pages}{2--16}.
\newblock


\bibitem[De~Nooy et~al\mbox{.}(2011)]%
        {de2011exploratory}
\bibfield{author}{\bibinfo{person}{Wouter De~Nooy}, \bibinfo{person}{Andrej
  Mrvar}, {and} \bibinfo{person}{Vladimir Batagelj}.}
  \bibinfo{year}{2011}\natexlab{}.
\newblock \bibinfo{booktitle}{\emph{Exploratory social network analysis with
  Pajek}}. Vol.~\bibinfo{volume}{27}.
\newblock \bibinfo{publisher}{Cambridge University Press}.
\newblock


\bibitem[Di~Cecco et~al\mbox{.}(2021)]%
        {di2021observing}
\bibfield{author}{\bibinfo{person}{Grace~J Di~Cecco}, \bibinfo{person}{Vijay
  Barve}, \bibinfo{person}{Michael~W Belitz}, \bibinfo{person}{Brian~J Stucky},
  \bibinfo{person}{Robert~P Guralnick}, {and} \bibinfo{person}{Allen~H
  Hurlbert}.} \bibinfo{year}{2021}\natexlab{}.
\newblock \showarticletitle{Observing the observers: How participants
  contribute data to iNaturalist and implications for biodiversity science}.
\newblock \bibinfo{journal}{\emph{BioScience}} \bibinfo{volume}{71},
  \bibinfo{number}{11} (\bibinfo{year}{2021}), \bibinfo{pages}{1179--1188}.
\newblock


\bibitem[Leskovec et~al\mbox{.}(2007)]%
        {leskovec_networks}
\bibfield{author}{\bibinfo{person}{Jure Leskovec}, \bibinfo{person}{Jon
  Kleinberg}, {and} \bibinfo{person}{Christos Faloutsos}.}
  \bibinfo{year}{2007}\natexlab{}.
\newblock \showarticletitle{Graph Evolution: Densification and Shrinking
  Diameters}.
\newblock \bibinfo{journal}{\emph{ACM Trans. Knowl. Discov. Data}}
  \bibinfo{volume}{1}, \bibinfo{number}{1} (\bibinfo{date}{mar}
  \bibinfo{year}{2007}), \bibinfo{pages}{2–es}.
\newblock
\showISSN{1556-4681}
\urldef\tempurl%
\url{https://doi.org/10.1145/1217299.1217301}
\showDOI{\tempurl}


\bibitem[Mazumdar and Thakker(2020)]%
        {fi12120210}
\bibfield{author}{\bibinfo{person}{Suvodeep Mazumdar} {and}
  \bibinfo{person}{Dhavalkumar Thakker}.} \bibinfo{year}{2020}\natexlab{}.
\newblock \showarticletitle{Citizen Science on Twitter: Using Data Analytics to
  Understand Conversations and Networks}.
\newblock \bibinfo{journal}{\emph{Future Internet}} \bibinfo{volume}{12},
  \bibinfo{number}{12} (\bibinfo{year}{2020}).
\newblock
\showISSN{1999-5903}
\urldef\tempurl%
\url{https://doi.org/10.3390/fi12120210}
\showDOI{\tempurl}


\bibitem[Newman(2006)]%
        {newman2006finding}
\bibfield{author}{\bibinfo{person}{Mark~EJ Newman}.}
  \bibinfo{year}{2006}\natexlab{}.
\newblock \showarticletitle{Finding community structure in networks using the
  eigenvectors of matrices}.
\newblock \bibinfo{journal}{\emph{Physical review E}} \bibinfo{volume}{74},
  \bibinfo{number}{3} (\bibinfo{year}{2006}), \bibinfo{pages}{036104}.
\newblock


\bibitem[Pelacho et~al\mbox{.}(2020)]%
        {citizen_science}
\bibfield{author}{\bibinfo{person}{M. Pelacho}, \bibinfo{person}{G. Ruiz},
  \bibinfo{person}{F. Sanz}, \bibinfo{person}{A. Tarancon}, {and}
  \bibinfo{person}{J. Clemente-Gallardo}.} \bibinfo{year}{2020}\natexlab{}.
\newblock \showarticletitle{Analysis of the evolution and collaboration
  networks of citizen scientific publications}.
\newblock \bibinfo{journal}{\emph{AI Magazine}} \bibinfo{volume}{126},
  \bibinfo{number}{3} (\bibinfo{date}{Oct.} \bibinfo{year}{2020}),
  \bibinfo{pages}{225}.
\newblock
\urldef\tempurl%
\url{https://doi.org/10.1007/s11192-020-03724-x}
\showDOI{\tempurl}


\bibitem[Perozzi et~al\mbox{.}(2014)]%
        {deepwalk}
\bibfield{author}{\bibinfo{person}{Bryan Perozzi}, \bibinfo{person}{Rami
  Al-Rfou}, {and} \bibinfo{person}{Steven Skiena}.}
  \bibinfo{year}{2014}\natexlab{}.
\newblock \showarticletitle{{DeepWalk}}. In
  \bibinfo{booktitle}{\emph{Proceedings of the 20th {ACM} {SIGKDD}
  international conference on Knowledge discovery and data mining}}.
  \bibinfo{publisher}{{ACM}}.
\newblock
\urldef\tempurl%
\url{https://doi.org/10.1145/2623330.2623732}
\showDOI{\tempurl}


\bibitem[Rossi and Ahmed(2015)]%
        {network_repository}
\bibfield{author}{\bibinfo{person}{Ryan~A. Rossi} {and}
  \bibinfo{person}{Nesreen~K. Ahmed}.} \bibinfo{year}{2015}\natexlab{}.
\newblock \showarticletitle{The Network Data Repository with Interactive Graph
  Analytics and Visualization}. In \bibinfo{booktitle}{\emph{AAAI}}.
\newblock
\urldef\tempurl%
\url{https://networkrepository.com}
\showURL{%
\tempurl}


\bibitem[Sen et~al\mbox{.}(2008)]%
        {pubmed_cora_citeseer}
\bibfield{author}{\bibinfo{person}{Prithviraj Sen}, \bibinfo{person}{Galileo
  Namata}, \bibinfo{person}{Mustafa Bilgic}, \bibinfo{person}{Lise Getoor},
  \bibinfo{person}{Brian Galligher}, {and} \bibinfo{person}{Tina Eliassi-Rad}.}
  \bibinfo{year}{2008}\natexlab{}.
\newblock \showarticletitle{Collective Classification in Network Data}.
\newblock \bibinfo{journal}{\emph{AI Magazine}} \bibinfo{volume}{29},
  \bibinfo{number}{3} (\bibinfo{date}{Sep.} \bibinfo{year}{2008}),
  \bibinfo{pages}{93}.
\newblock
\urldef\tempurl%
\url{https://doi.org/10.1609/aimag.v29i3.2157}
\showDOI{\tempurl}


\bibitem[Wikipedia(2022a)]%
        {wiki:Agaricomycetes}
\bibfield{author}{\bibinfo{person}{Wikipedia}.}
  \bibinfo{year}{2022}\natexlab{a}.
\newblock \bibinfo{title}{{Agaricomycetes} --- {W}ikipedia{,} The Free
  Encyclopedia}.
\newblock
  \bibinfo{howpublished}{\url{http://en.wikipedia.org/w/index.php?title=Agaricomycetes&oldid=1113445436}}.
\newblock
\newblock
\shownote{[Online; accessed 05-December-2022]}.


\bibitem[Wikipedia(2022b)]%
        {wiki:Alloclavaria_purpurea}
\bibfield{author}{\bibinfo{person}{Wikipedia}.}
  \bibinfo{year}{2022}\natexlab{b}.
\newblock \bibinfo{title}{{Alloclavaria purpurea} --- {W}ikipedia{,} The Free
  Encyclopedia}.
\newblock
  \bibinfo{howpublished}{\url{http://en.wikipedia.org/w/index.php?title=Alloclavaria\%20purpurea&oldid=1051410601}}.
\newblock
\newblock
\shownote{[Online; accessed 05-December-2022]}.


\bibitem[Wikipedia(2022c)]%
        {wiki:American_black_bear}
\bibfield{author}{\bibinfo{person}{Wikipedia}.}
  \bibinfo{year}{2022}\natexlab{c}.
\newblock \bibinfo{title}{{American black bear} --- {W}ikipedia{,} The Free
  Encyclopedia}.
\newblock
  \bibinfo{howpublished}{\url{http://en.wikipedia.org/w/index.php?title=American\%20black\%20bear&oldid=1121916752}}.
\newblock
\newblock
\shownote{[Online; accessed 05-December-2022]}.


\bibitem[Wikipedia(2022d)]%
        {wiki:Aquilegia_flavescens}
\bibfield{author}{\bibinfo{person}{Wikipedia}.}
  \bibinfo{year}{2022}\natexlab{d}.
\newblock \bibinfo{title}{{Aquilegia flavescens} --- {W}ikipedia{,} The Free
  Encyclopedia}.
\newblock
  \bibinfo{howpublished}{\url{http://en.wikipedia.org/w/index.php?title=Aquilegia\%20flavescens&oldid=1094178920}}.
\newblock
\newblock
\shownote{[Online; accessed 05-December-2022]}.


\bibitem[Wikipedia(2022e)]%
        {wiki:Bighorn_sheep}
\bibfield{author}{\bibinfo{person}{Wikipedia}.}
  \bibinfo{year}{2022}\natexlab{e}.
\newblock \bibinfo{title}{{Bighorn sheep} --- {W}ikipedia{,} The Free
  Encyclopedia}.
\newblock
  \bibinfo{howpublished}{\url{http://en.wikipedia.org/w/index.php?title=Bighorn\%20sheep&oldid=1125781182}}.
\newblock
\newblock
\shownote{[Online; accessed 05-December-2022]}.


\bibitem[Wikipedia(2022f)]%
        {wiki:Bird}
\bibfield{author}{\bibinfo{person}{Wikipedia}.}
  \bibinfo{year}{2022}\natexlab{f}.
\newblock \bibinfo{title}{{Bird} --- {W}ikipedia{,} The Free Encyclopedia}.
\newblock
  \bibinfo{howpublished}{\url{http://en.wikipedia.org/w/index.php?title=Bird&oldid=1124903835}}.
\newblock
\newblock
\shownote{[Online; accessed 05-December-2022]}.


\bibitem[Wikipedia(2022g)]%
        {wiki:Canada_jay}
\bibfield{author}{\bibinfo{person}{Wikipedia}.}
  \bibinfo{year}{2022}\natexlab{g}.
\newblock \bibinfo{title}{{Canada jay} --- {W}ikipedia{,} The Free
  Encyclopedia}.
\newblock
  \bibinfo{howpublished}{\url{http://en.wikipedia.org/w/index.php?title=Canada\%20jay&oldid=1122059444}}.
\newblock
\newblock
\shownote{[Online; accessed 05-December-2022]}.


\bibitem[Wikipedia(2022h)]%
        {wiki:Chamaenerion_angustifolium}
\bibfield{author}{\bibinfo{person}{Wikipedia}.}
  \bibinfo{year}{2022}\natexlab{h}.
\newblock \bibinfo{title}{{Chamaenerion angustifolium} --- {W}ikipedia{,} The
  Free Encyclopedia}.
\newblock
  \bibinfo{howpublished}{\url{http://en.wikipedia.org/w/index.php?title=Chamaenerion\%20angustifolium&oldid=1125480878}}.
\newblock
\newblock
\shownote{[Online; accessed 05-December-2022]}.


\bibitem[Wikipedia(2022i)]%
        {wiki:Clark's_nutcracker}
\bibfield{author}{\bibinfo{person}{Wikipedia}.}
  \bibinfo{year}{2022}\natexlab{i}.
\newblock \bibinfo{title}{{Clark's nutcracker} --- {W}ikipedia{,} The Free
  Encyclopedia}.
\newblock
  \bibinfo{howpublished}{\url{http://en.wikipedia.org/w/index.php?title=Clark's\%20nutcracker&oldid=1122867502}}.
\newblock
\newblock
\shownote{[Online; accessed 05-December-2022]}.


\bibitem[Wikipedia(2022j)]%
        {wiki:Columbian_ground_squirrel}
\bibfield{author}{\bibinfo{person}{Wikipedia}.}
  \bibinfo{year}{2022}\natexlab{j}.
\newblock \bibinfo{title}{{Columbian ground squirrel} --- {W}ikipedia{,} The
  Free Encyclopedia}.
\newblock
  \bibinfo{howpublished}{\url{http://en.wikipedia.org/w/index.php?title=Columbian\%20ground\%20squirrel&oldid=1124139650}}.
\newblock
\newblock
\shownote{[Online; accessed 05-December-2022]}.


\bibitem[Wikipedia(2022k)]%
        {wiki:Coprinus_comatus}
\bibfield{author}{\bibinfo{person}{Wikipedia}.}
  \bibinfo{year}{2022}\natexlab{k}.
\newblock \bibinfo{title}{{Coprinus comatus} --- {W}ikipedia{,} The Free
  Encyclopedia}.
\newblock
  \bibinfo{howpublished}{\url{http://en.wikipedia.org/w/index.php?title=Coprinus\%20comatus&oldid=1106032411}}.
\newblock
\newblock
\shownote{[Online; accessed 05-December-2022]}.


\bibitem[Wikipedia(2022l)]%
        {wiki:Golden-mantled_ground_squirrel}
\bibfield{author}{\bibinfo{person}{Wikipedia}.}
  \bibinfo{year}{2022}\natexlab{l}.
\newblock \bibinfo{title}{{Golden-mantled ground squirrel} --- {W}ikipedia{,}
  The Free Encyclopedia}.
\newblock
  \bibinfo{howpublished}{\url{http://en.wikipedia.org/w/index.php?title=Golden-mantled\%20ground\%20squirrel&oldid=1121777526}}.
\newblock
\newblock
\shownote{[Online; accessed 05-December-2022]}.


\bibitem[Wikipedia(2022m)]%
        {wiki:Hydnellum_caeruleum}
\bibfield{author}{\bibinfo{person}{Wikipedia}.}
  \bibinfo{year}{2022}\natexlab{m}.
\newblock \bibinfo{title}{{Hydnellum caeruleum} --- {W}ikipedia{,} The Free
  Encyclopedia}.
\newblock
  \bibinfo{howpublished}{\url{http://en.wikipedia.org/w/index.php?title=Hydnellum\%20caeruleum&oldid=1110709335}}.
\newblock
\newblock
\shownote{[Online; accessed 05-December-2022]}.


\bibitem[Wikipedia(2022n)]%
        {wiki:INaturalist}
\bibfield{author}{\bibinfo{person}{Wikipedia}.}
  \bibinfo{year}{2022}\natexlab{n}.
\newblock \bibinfo{title}{{INaturalist} --- {W}ikipedia{,} The Free
  Encyclopedia}.
\newblock
  \bibinfo{howpublished}{\url{http://en.wikipedia.org/w/index.php?title=INaturalist&oldid=1122751314}}.
\newblock
\newblock
\shownote{[Online; accessed 05-December-2022]}.


\bibitem[Wikipedia(2022o)]%
        {wiki:Liliopsida}
\bibfield{author}{\bibinfo{person}{Wikipedia}.}
  \bibinfo{year}{2022}\natexlab{o}.
\newblock \bibinfo{title}{{Liliopsida} --- {W}ikipedia{,} The Free
  Encyclopedia}.
\newblock
  \bibinfo{howpublished}{\url{http://en.wikipedia.org/w/index.php?title=Liliopsida&oldid=1062713963}}.
\newblock
\newblock
\shownote{[Online; accessed 05-December-2022]}.


\bibitem[Wikipedia(2022p)]%
        {wiki:Magnoliopsida}
\bibfield{author}{\bibinfo{person}{Wikipedia}.}
  \bibinfo{year}{2022}\natexlab{p}.
\newblock \bibinfo{title}{{Magnoliopsida} --- {W}ikipedia{,} The Free
  Encyclopedia}.
\newblock
  \bibinfo{howpublished}{\url{http://en.wikipedia.org/w/index.php?title=Magnoliopsida&oldid=1125274045}}.
\newblock
\newblock
\shownote{[Online; accessed 05-December-2022]}.


\bibitem[Wikipedia(2022q)]%
        {wiki:Phellinus_tremulae}
\bibfield{author}{\bibinfo{person}{Wikipedia}.}
  \bibinfo{year}{2022}\natexlab{q}.
\newblock \bibinfo{title}{{Phellinus tremulae} --- {W}ikipedia{,} The Free
  Encyclopedia}.
\newblock
  \bibinfo{howpublished}{\url{http://en.wikipedia.org/w/index.php?title=Phellinus\%20tremulae&oldid=1008806146}}.
\newblock
\newblock
\shownote{[Online; accessed 05-December-2022]}.


\bibitem[Wikipedia(2022r)]%
        {wiki:Red_deer}
\bibfield{author}{\bibinfo{person}{Wikipedia}.}
  \bibinfo{year}{2022}\natexlab{r}.
\newblock \bibinfo{title}{{Red deer} --- {W}ikipedia{,} The Free Encyclopedia}.
\newblock
  \bibinfo{howpublished}{\url{http://en.wikipedia.org/w/index.php?title=Red\%20deer&oldid=1125200239}}.
\newblock
\newblock
\shownote{[Online; accessed 05-December-2022]}.


\bibitem[Wikipedia(2022s)]%
        {wiki:Sarcodon_imbricatus}
\bibfield{author}{\bibinfo{person}{Wikipedia}.}
  \bibinfo{year}{2022}\natexlab{s}.
\newblock \bibinfo{title}{{Sarcodon imbricatus} --- {W}ikipedia{,} The Free
  Encyclopedia}.
\newblock
  \bibinfo{howpublished}{\url{http://en.wikipedia.org/w/index.php?title=Sarcodon\%20imbricatus&oldid=1121756904}}.
\newblock
\newblock
\shownote{[Online; accessed 05-December-2022]}.


\bibitem[Wilson et~al\mbox{.}(2020)]%
        {bumblebee}
\bibfield{author}{\bibinfo{person}{Joseph~S. Wilson}, \bibinfo{person}{Aaron~D.
  Pan}, \bibinfo{person}{David Emmanuel~M. General}, {and}
  \bibinfo{person}{Jonathan~B. Koch}.} \bibinfo{year}{2020}\natexlab{}.
\newblock \showarticletitle{More eyes on the prize: an observation of a very
  rare, threatened species of Philippine Bumble bee, Bombus irisanensis, on
  iNaturalist and the importance of citizen science in conservation biology}.
\newblock \bibinfo{journal}{\emph{Journal of Insect Conservation}}
  \bibinfo{volume}{24}, \bibinfo{number}{4} (\bibinfo{date}{01 Aug}
  \bibinfo{year}{2020}), \bibinfo{pages}{727--729}.
\newblock
\showISSN{1572-9753}
\urldef\tempurl%
\url{https://doi.org/10.1007/s10841-020-00233-3}
\showDOI{\tempurl}


\end{thebibliography}

\end{document}